\documentclass[conference]{IEEEtran}
\IEEEoverridecommandlockouts
\usepackage{graphicx}
\usepackage{float}
\usepackage{subfigure}
\usepackage{amssymb}
\usepackage{dsfont}
\usepackage{epstopdf}
\usepackage{booktabs}
\usepackage{multirow}
\usepackage[normalem]{ulem}
\useunder{\uline}{\ul}{}
\usepackage{cite}
\usepackage{amsmath,amssymb,amsfonts}
\usepackage{algorithmic}
\usepackage{graphicx}
\usepackage{textcomp}
\usepackage{xcolor}
\usepackage[letterpaper,top=54pt,bottom=54pt,left=54pt,right=54pt]{geometry}
\usepackage[T1]{fontenc}
\usepackage{aecompl}

\def\BibTeX{{\rm B\kern-.05em{\sc i\kern-.025em b}\kern-.08em
    T\kern-.1667em\lower.7ex\hbox{E}\kern-.125emX}}
\usepackage{hyperref} % 宏包
\hypersetup{
    colorlinks=true,
    linkcolor=black,
    anchorcolor=black,
    citecolor=black,
    urlcolor=black
}  % 对应设置
\begin{document}

% 字体命令
%求求收了吧，这篇论文
\title{Topology-aware Debiased Self-supervised Graph Learning for Recommendation}

\author{\IEEEauthorblockN{1\textsuperscript{st} Lei Han}
\IEEEauthorblockA{\textit{School of Computer Science and} \\
\textit{Engineering} \\
\textit{Nanjing University of Science and}\\
\textit{Technology}\\
Nanjing, China \\
hanl@njust.edu.cn}
\and
\IEEEauthorblockN{2\textsuperscript{nd} Hui Yan}
\IEEEauthorblockA{\textit{School of Computer Science and} \\
\textit{Engineering} \\
\textit{Nanjing University of Science and}\\
\textit{Technology}\\
Nanjing, China \\
yanhui@njust.edu.cn}
\and
\IEEEauthorblockN{3\textsuperscript{rd} Zhicheng Qiao}
\IEEEauthorblockA{\textit{School of Computer Science and} \\
\textit{Engineering} \\
\textit{Nanjing University of Science and}\\
\textit{Technology}\\
Nanjing, China \\
zchengqiao@njust.edu.cn}
}

\maketitle
\begin{abstract}
In recommendation, graph-based Collaborative Filtering (CF) methods mitigate the data sparsity by introducing Graph Contrastive Learning (GCL). However, the random negative sampling strategy in these GCL-based CF models neglects the semantic structure of users (items), which not only introduces false negatives (negatives that are similar to anchor user (item)) but also ignores the potential positive samples. To tackle the above issues, we propose Topology-aware Debiased Self-supervised Graph Learning (TDSGL) for recommendation, which constructs contrastive pairs according to the semantic similarity between users (items). Specifically, since the original user-item interaction data commendably reflects the purchasing intent of users and certain characteristics of items, we calculate the semantic similarity between users (items) on interaction data. Then, given a user (item), we construct its negative pairs by selecting users (items) which embed different semantic structures to ensure the semantic difference between the given user (item) and its negatives. Moreover, for a user (item), we design a feature extraction module that converts other semantically similar users (items) into an auxiliary positive sample to acquire a more informative representation. Experimental results show that the proposed model outperforms the state-of-the-art models significantly on three public datasets. Our model implementation codes are available at \url{https://github.com/malajikuai/TDSGL}.
\end{abstract}


\section{Introduction}
\begin{figure}[t]
\centering
\subfigure[user-item interactions graph]{
\label{Fig1.sub.1}
\includegraphics[width=3.5cm, height = 4.5cm]{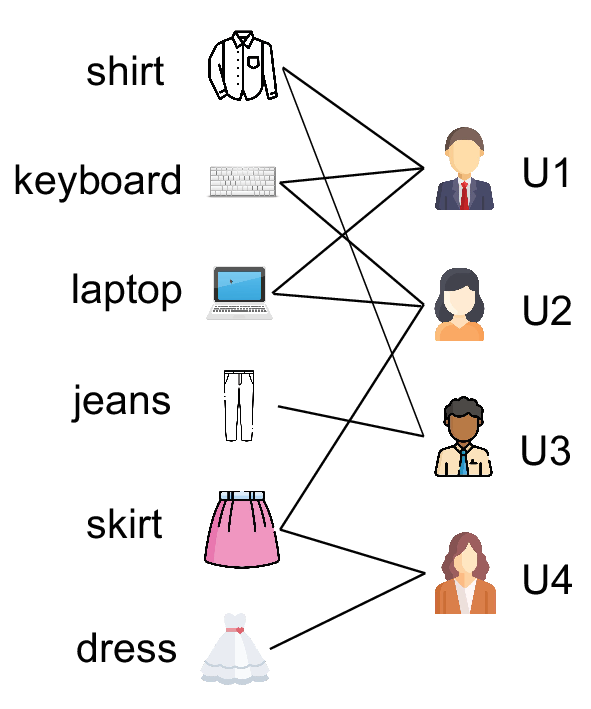}}
\subfigure[negative samples construction]{
\label{Fig1.sub.2}
\includegraphics[width=4.5cm, height = 4.5cm]{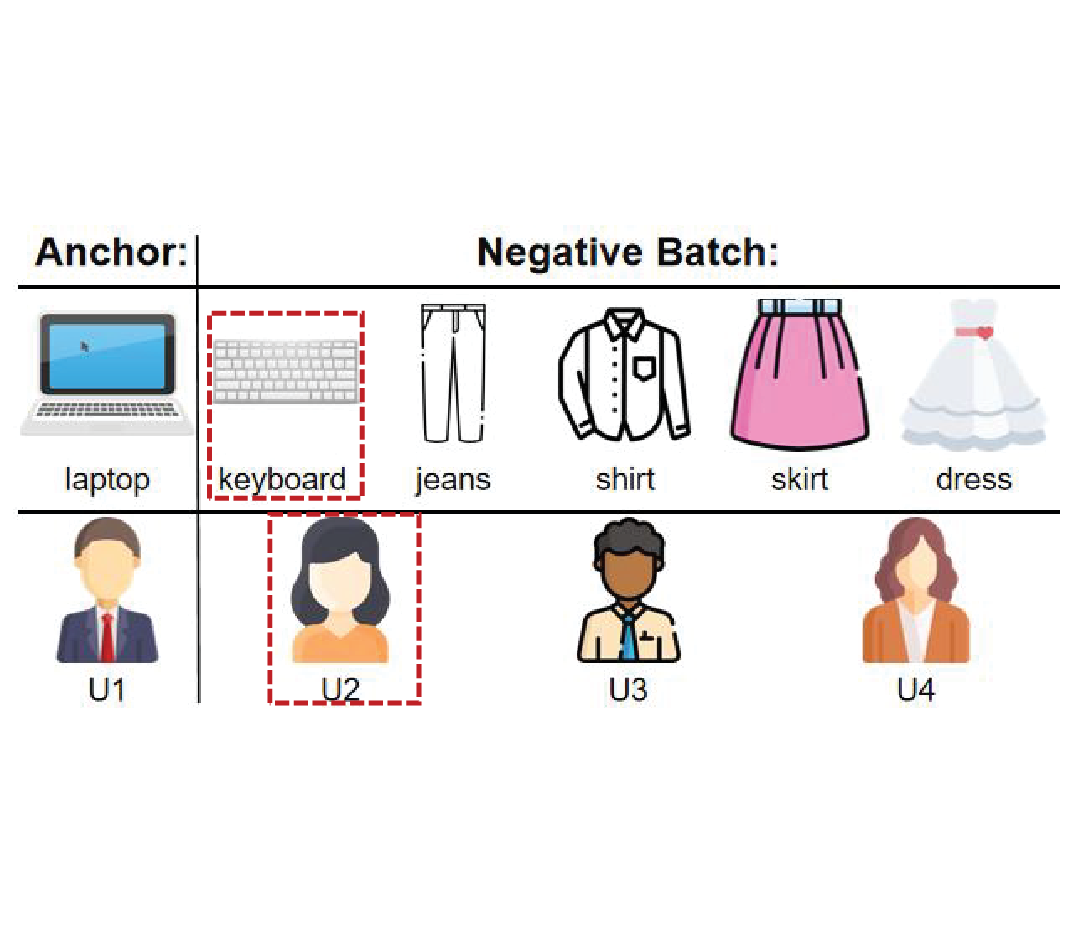}}
\caption{Examples of user-item interactions and negative samples in contrastive learning of SGL.
(a) is user-item interactions’ bipartite graph, (b) describes the negative samples to certain user (or item)
which are uniformly chosen from the rest users (or items).
}  % 图片标题
\label{Fig 1}    % 标签，用来引用
\end{figure}

% This document is a model and instructions for \LaTeX.
% Please observe the conference page limits. 
%第一段：推荐系统和CF
In such an era of information, it is necessary to effectively extract informative representations from the previous interactions between users and items for Recommender Systems (RS) \cite{huang2019online,covington2016deep,li2020lstm}. Collaborative Filtering (CF) is one of the most successful and popular methods in recommendation, which holds the following assumption: users with similar interests in the past will express common interests in the future \cite{wu2022graph}. To take advantage of the natural graph structure of user-item interaction data (illustrated in Fig. \ref{Fig1.sub.1}), Graph Convolutional Network (GCN) based CF algorithms have been developed to improve the performance of RS by capturing high-order connectivity among users and items. Representative works include Pinsage \cite{ying2018graph}, NGCF \cite{wang2019neural}, LightGCN \cite{he2020lightgcn}, LRGCCF \cite{chen2020revisiting}, and DGCF \cite{wang2020disentangled}. 

%第二段：GCN的不足以及CL的引入（主要：数据稀缺性）
Although the above graph neural collaborative filtering algorithms have improved the recommendation accuracy, they are still vulnerable to the sparse user-item interaction data because of Graph Neural Networks (GNN) characteristics. To reduce the influence of data sparsity, \cite{wu2021self,yao2021self} introduce Graph Contrastive Learning (GCL) as Self-supervised Learning (SSL) task and leverage multi-task strategy to optimize the conventional recommendation task and auxiliary SSL task jointly. In general, GCL constructs multiple views via stochastic augmentations of the input data and then learning representations by contrasting positive samples against negative samples (negatives) \cite{zhu2021empirical}.

% \begin{figure}[t]
% \centering
% \includegraphics[height=5cm,width=8cm]{1_eps.eps}  % 图片路径
% \caption{ Examples of user-item interactions and negative samples in contrastive learning of SGL.
% (a) is user-item interactions’ bipartite graph, (b) describes the negative samples to certain user (or item)
% which are uniformly choosed from the rest users and items.}  % 图片标题
% \label{Fig 1}    % 标签，用来引用
% \end{figure}

%第三段对比学习的引入以及其缺陷
The efficacy of Contrastive Learning (CL) is heavily reliant on the selection of positive and negative samples \cite{robinson2020contrastive}. The positive samples in \cite{wu2021self,yao2021self} are defined as the samples from the same user (item), while the negative samples are from all different users (items). However, the method of negative sampling without discrimination among different users (items), as shown in Fig. \ref{Fig1.sub.2}, may introduce sampling bias, resulting in significant performance degradation \cite{chuang2020debiased}. For example, it is inappropriate to consider the keyboard as a negative sample of the laptop, as these two items often interact with many common users. Similarly, U1 and U2 share similar purchasing interests, rendering U1 a potential positive sample of U2. Nonetheless, existing GCL-based CF methods fail to identify such positive samples. Therefore, we argue it is unreasonable to use the original sampling strategy because it would result in sampling bias and ignore the potential positive samples.

%第四段：直接说解决方法
Many existing works \cite{kaya2019deep,chuang2020debiased,xu2022negative}, both theoretically and practically validate that removing false negatives improves the performance of CL, but several studies \cite{zhu2021empirical,xia2022progcl} report that adopting these negative mining techniques based on feature brings minor improvement in GCL at the same time (we report the similar phenomenon in GCL-based CF in Table \ref{tab:Comparance with variants}). As the interaction behavior between users and items inherently reflects the user's purchasing interests, the overlap degree of purchased items among users can be considered a form of similarity (the same holds for items). This motivates our topological approach to optimizing the selection of positive and negative samples in GCL-based CF. 

To mitigate the sampling bias issue and further utilize the potential positive samples, we propose Topology-aware Debiased Self-supervised Graph Learning for Recommendation (TDSGL), a new framework that calculates the semantic similarity between each pair of users (items) based on the user-item interactions to construct positive samples and negative samples. Specifically, given a query, in order to reduce the influence of false negatives, we perform negative sampling via selecting the users (items) with low similarity to it. Moreover, we devise a feature extraction module, which extracts the semantic information of potential positive samples, to solve the imbalanced potential positive samples distribution problem. In summary, we make the following contributions:
\begin{itemize}
\item We propose a simple yet effective method that can classify the negative samples into true and false ones on interaction data and decrease the false negatives in negative pairs.
\item In practice, we think that false negatives are actually potential positive samples and devise a GCN-based feature extraction module to solve the imbalanced potential positive samples distribution problem.
\item Combining both technical contributions into a single model, TDSGL outperforms the state-of-the-art methods in yielding better performance on three benchmark datasets.
\end{itemize}

\section{RELATED WORK}

\subsection{Graph-based Collaborative Filtering}\label{GCF} 
Unlike traditional Matrix Factorization-based methods \cite{liu2015boosting,liu2017learning,koren2009matrix}, graph-based collaborative filtering integrates multi-hop neighbors into node representation learning to enhance model performance \cite{gao2021graph}. Specifically, NGCF \cite{wang2019neural} and PinSage \cite{ying2018graph} successfully utilize GCN to capture the high-hop neighbors’ information. Furthermore, LR-GCCF \cite{chen2020revisiting} and LightGCN \cite{he2020lightgcn} share similar ideas to simplify the heavy networks of GCN to enhance the performance. DGCF \cite{wang2020disentangled} aims to obtain intent-aware representations via modeling diverse user-item interactions. Since these models still suffer from sparse and noisy interaction data, \cite{wu2021self,yao2021self} introduce GCL into recommendation as a self-supervised task and achieve desirable performance. In general, these models perform data augmentations and contrastive learning these two steps, then the classical supervised task of recommendation is combined with the auxiliary self-supervised task in the optimization step. Although GCL-based CF algorithms enhance recommendation accuracy, they lack consideration for constructing more appropriate contrastive pairs tailored for the recommendation task.

\subsection{Sampling bias in Graph Contrastive Learning}\label{Sampling bias} 
There are several studies proposed to solve the sampling bias in GCL. Zhu et al. \cite{zhu2021empirical} observe that existing hard negative mining strategies based on calculating embedding similarities bring limited improvements to GCL. Xia et al. \cite{xia2022progcl} explain why existing negative mining techniques can not work well in GCL and utilizes the beta mixture model to estimate the probability of a negative sample being true one relative to a specific anchor. Zhao et al. \cite{zhao2021graph} utilize the clustering pseudo labels to alleviate the issue of the false negative, but it suffers from heavy computational overhead and will degrade the performance when confronted with multi-class datasets. Lin et al. \cite{lin2022prototypical} add a constraint to assure a reasonable clusters assignment for graph-level contrastive learning and emphasize the negatives whose clusters embed a moderate distance to anchor. However, for node-level contrastive learning work in RS, it is difficult to know precisely how many clusters there are and hard to apply the method in \cite{xia2022progcl} limited by datasets.

\section{METHODOLOGY} 

\begin{figure*}[t]
\centering
\subfigure[The overall system framework of TDSGL]{
\label{Fig2.sub.1}
\includegraphics[width=9cm, height = 5.0cm]{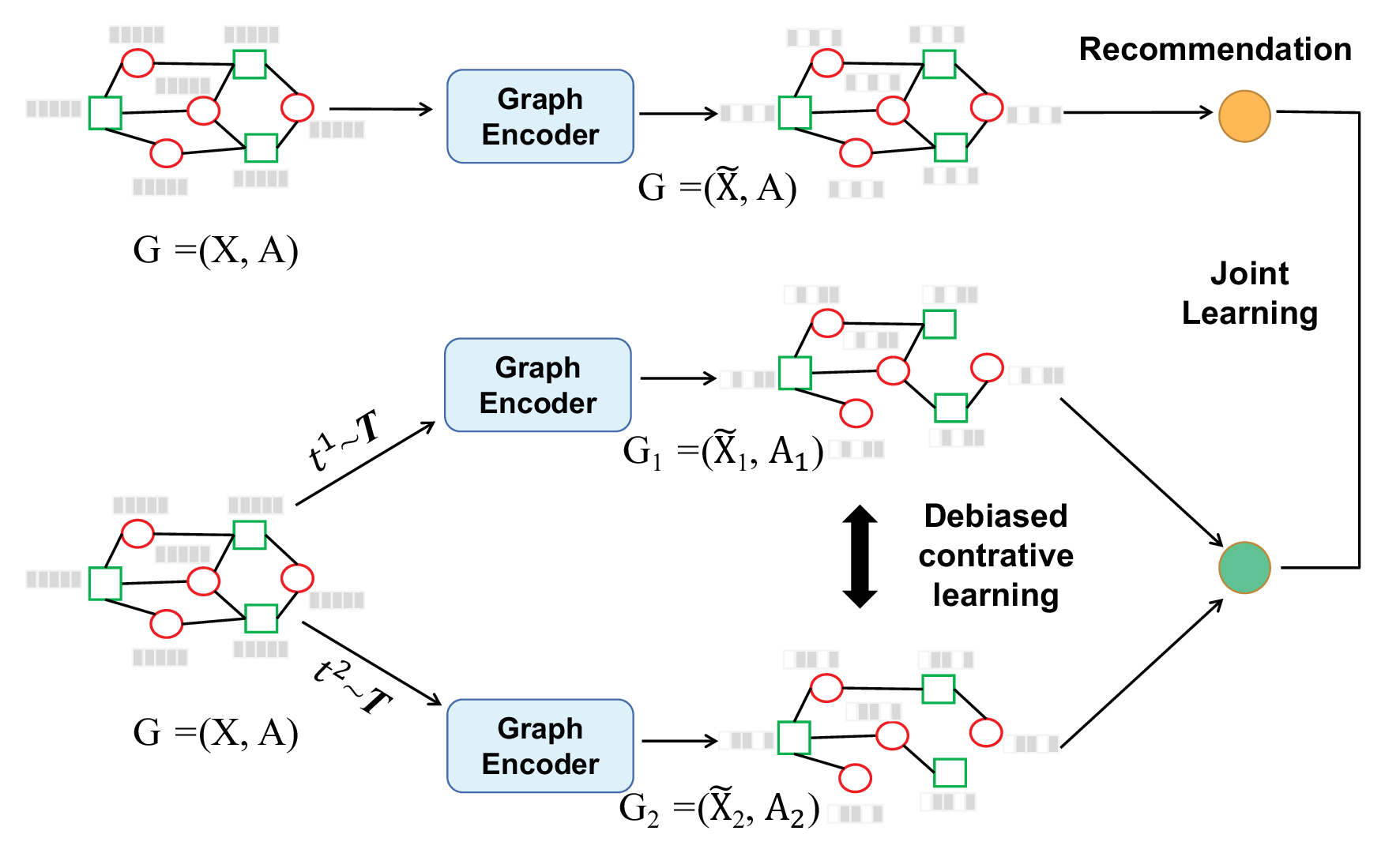}}
\subfigure[Details of contrastive learning in TDSGL]{
\label{Fig2.sub.2}
\includegraphics[width=8cm, height = 5.0cm]{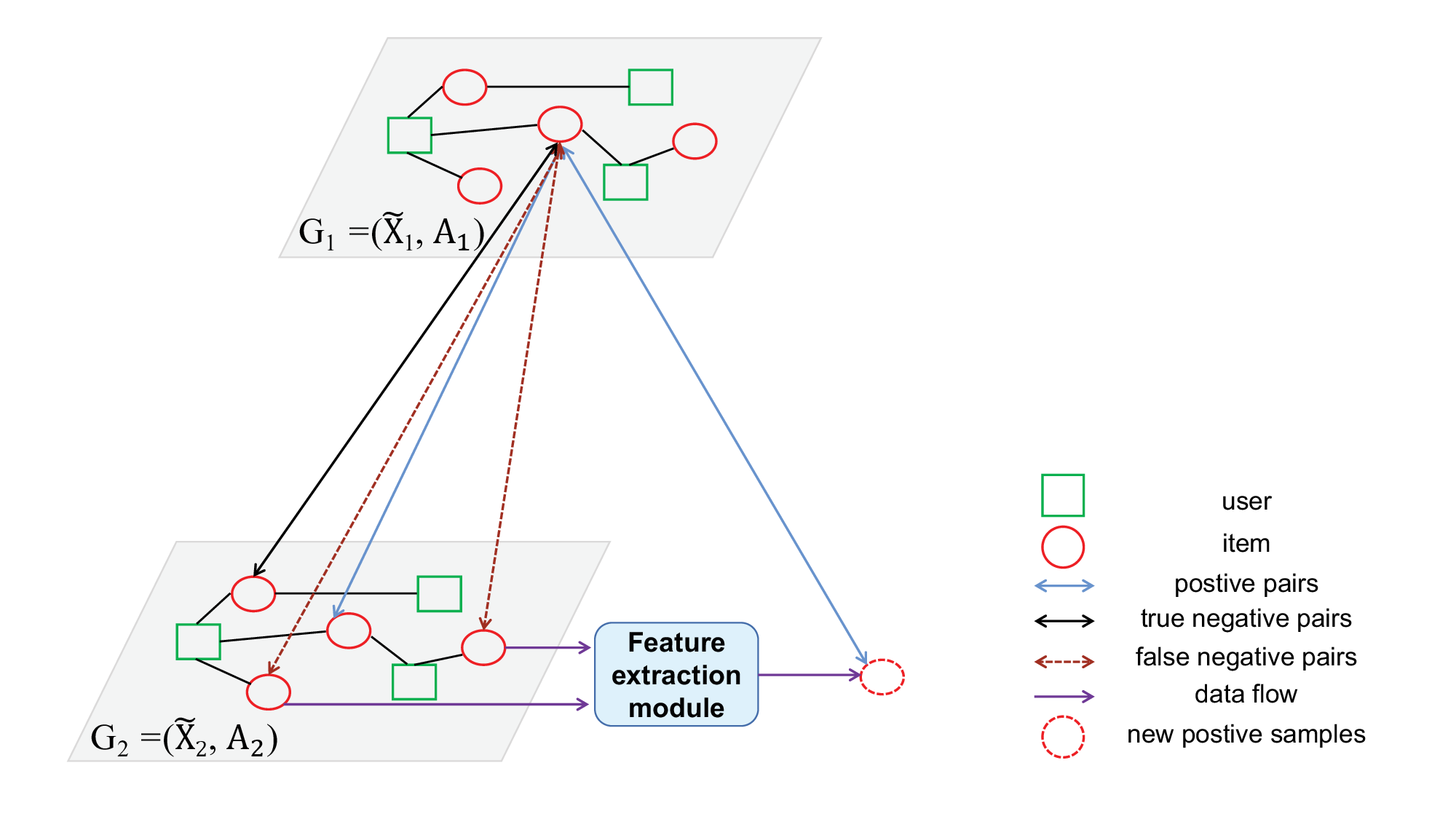}}
\caption{Framework (a) of the proposed TDSGL, the upper layer illustrates the working flow of the recommendation task while the bottom layer shows the working flows of debiased contrastive learning task with data augmentation \textbf{\textit{T}} on graph structure. (b) Compared to conventional GCL-based CF, TDSGL not only effectively decreases the false negatives in the negative sampling process (the red dash line means that TDSGL no longer treats these samples as negatives), but also devises a GCN-based feature extraction module to converts them into auxiliary positive samples.
}  % 图片标题
\label{Fig 2}    % 标签，用来引用
\end{figure*}

In this section, we first review the background and problem settings of traditional GCN-based collaborative filtering in Section \ref{3.1}. To reduce the false negatives of existing contrastive learning methods, we propose our topological method of false negatives exploration in Section \ref{3.2}. Furthermore, we devise a feature extraction module to convert these false negatives into positive samples in Section \ref{3.3}. The Fig. \ref{Fig 2} illustrates the working flow of TDSGL. Lastly, in Section \ref{3.4}, our debiased GCL is combined with classical GCN in a multi-task learning manner.

\subsection{Preliminaries}\label{3.1}
\paragraph{\textbf{Notations}}
Defining $\mathcal{G = (V, E)}$ be a given interaction graph in which nodes set $\mathcal{V}$ consists of user nodes $u\in\mathcal{U}$ and item nodes $i\in\mathcal{I}$, the edge set $\mathcal{E}$ depicts the interactions between users and items. The number of users and items are denoted by $N_U$ and $ N_I$, then we denote $N=N_U + N_I$. The adjacency matrix is constructed from the user-item interaction matrix $\mathbf{R} \in \mathbb{R}^{N_U \times N_I}$ and denoted as $\mathbf{A} \in \mathbb{R}^{N \times N}$, the diagonal degree matrix is denoted as $\mathbf{D} \in \mathbb{R}^{N \times N}$. In recommender system, we always randomly initialize an embedding matrix $\mathbf{X}^{(0)} \in \mathbb{R}^{N \times F}$ to represent users’ and items’ latent features, where $F$ is the dimension of features. In contrastive learning, we utilize data augmentation \textbf{\textit{T}}, including edge dropout, node dropout, and mask, to generate two views $\mathrm{G_1}$ and $\mathrm{G_2}$. In the following discussion, we mainly use the matrix form notation of each model. 

\paragraph{\textbf{Graph-based collaborative filtering}}
Applying the propagation and prediction function on the interaction data, Graph-based CF obtains informative user and item representation. Following LightGCN, we discard the nonlinear activation and feature transformation in the propagation as:
\begin{equation}\label{GCN_eq}
\mathbf{X}^{(l+1)} = \hat{\mathbf{D}}^{-1/2}\hat{\mathbf{A}}\hat{\mathbf{D}}^{-1/2}\mathbf{X}^{(l)}
\end{equation}
\begin{equation}\label{finalGCN_eq}
\mathbf{X} = \frac{1}{L+1}\sum_{l=0}^{L}\mathbf{X}^{(l)}
\end{equation}
where ${\mathbf{X}^{(l+1)}}$ is the the hidden embedding at (${l + 1}$)th layer, $\hat{\mathbf{A}}$ is the adjacency matrix $\mathbf{A}$ with self-loop and $\hat{\mathbf{D}}$ is the diagonal degree matrix of $\hat{\mathbf{A}}$. After propagating with $L$ layers, we adopt the weighted sum function to combine the representations of all layers to obtain the final representations. 

In the prediction step, we adopt Bayesian Personalized Ranking (BPR) loss\cite{rendle2012bpr}, and the objective function of BPR loss is as follows:
\begin{equation}\label{rec_eq}
    \mathcal{L}_{rec} = \sum_{(u,i,j) \in \mathcal{O}} -log\  \sigma(f(x_u,x_i) - f(x_u,x_j)) 
\end{equation}
where $\mathcal{O}$ is the set of training instances that $u$ interacted with $i$ and didn't interact with $j$, $\sigma $ is a nonlinear activate function, $x_u$, $x_i$ and $x_j$ are the final embeddings of user $u$, item $i$ and item $j$, $f(\cdot)$ is the prediction score calculation rule, and here we just simply take the inner product.

\subsection{False negatives exploration}\label{3.2}
GCL-based recommender system always utilizes data augmentations on input data to generate multiple views and learns representations by contrasting positive samples against negative samples of users (items). However, the negative sampling strategy introduces false negatives as it ignores the semantic similarity between users (items). Given a user (item), we aim to distinguish the false negatives from all negatives. Because the original behavior history shows users' preferences and items' certain characteristics, we construct user-user and item-item two co-occurrence matrices based on the original interaction matrix $\mathbf{R}$. These two matrices can commendably reflect the similarity of users' purchasing intent and items' certain characteristics, the matrices are constructed as follows:
\begin{equation}\label{matrix_eq}
    \mathbf{P}_{user} = \mathbf{R} \times \mathbf{R}^T,\ \mathbf{P}_{item} = \mathbf{R}^{T} \times \mathbf{R}
\end{equation}
where $\mathbf{P}_{user} \in \mathbb{R}^{N_U \times N_U}$ and $\mathbf{P}_{item} \in \mathbb{R}^{N_I \times N_I}$, each entity in matrix $\mathbf{P}_{user}$ represents the purchasing intent similarity between every two users, this rule can also be applied to $\mathbf{P}_{item}$. Furthermore, we define a hyperparameter $\beta$ that represents the pre-defined threshold value to filter those entities with low similarity in co-occurrence matrices: 
\begin{equation}\label{refine_eq}
\resizebox{.9\hsize}{!}{$\mathbf{M}_{user}(u,v) = \mathbb{I}(\mathbf{P}_{user}(u,v) < \beta),\ \mathbf{M}_{item}(i,j) = \mathbb{I}(\mathbf{P}_{item}(i,j) < \beta)$}
\end{equation}
where $\mathbb{I}(\cdot)$ is a binary indicator function returning 1 when the condition is true, otherwise returning 0, and $\mathbf{M}_{user}(u,v) = 0$ meaning $u,v$ share the same purchasing interest. Formally, we follow SimCLR\cite{gutmann2010noise} and adopt the contrastive loss, InfoNCE\cite{chen2020simple}, then we acquire our debiased contrastive learning loss of the user side:
\begin{equation}\label{debiased_eq}
\resizebox{.9\hsize}{!}{$\mathcal{L}_{dbs\_user} = \sum_{u \in \mathcal{U}} -log \frac{exp(s(x_u^{'}, x_u^{''})/\tau)}{\sum_{v\in \mathcal{U}} \mathbf{M}_{user}(u,v) \cdot exp(s(x_u^{'},x_v^{''})/\tau)}$}
\end{equation}
where $s(\cdot)$ denotes the cosine similarity function and $\tau$ is the temperature hyper-parameter of softmax, $x^{'}$ and $x^{''}$ are node representations learned from two different graph augmentations \textbf{\textit{T}}. Analogously, we obtain the debiased contrastive loss of the item side $\mathcal{L}_{dbs\_item}$.

\subsection{Feature extraction module}\label{3.3}
Based on the presentation in Section \ref{3.2}, we can obtain the distribution of false negatives for each user (item). Since false negatives and users (items) share similar semantic information, we argue that it is not appropriate to remove them among negatives simply. On the contrary, these false negatives should be treated as potential positive samples. However, each user (item) has a different number of false negatives (some are more, and some are less). In order to solve the imbalanced false negatives distribution issue, we devise a GCN-based feature extraction module that treats $\mathbf{M}_{user}$ and $\mathbf{M}_{item}$ as adjacency matrices, and employ one-layer GCN to effectively extract the feature of false negatives corresponding to each user (item). The feature extraction can be written as follows:
\begin{equation}\label{u2u_eq}
    \mathbf{X}^{M\_u} = \mathbf{D}_{M\_user}^{-1/2}\mathbf{M}_{M\_user}\mathbf{D}_{M\_user}^{-1/2}\mathbf{X}_{user}^{(0)}
\end{equation}
\begin{equation}\label{i2i_eq}
    \mathbf{X}^{M\_i} = \mathbf{D}_{M\_item}^{-1/2}\mathbf{M}_{M\_item}\mathbf{D}_{M\_item}^{-1/2}\mathbf{X}_{item}^{(0)}
\end{equation}
where $\mathbf{X}^{M\_u} \in \mathbb{R}^{N_{U} \times F}$ and $\mathbf{X}^{M\_i} \in \mathbb{R}^{N_{I} \times F}$, each column in $\mathbf{X}^{M\_u}$ (or $\mathbf{X}^{M\_i}$) represents the feature of false negatives corresponding to a certain user (or item), we treat these features as positive samples. The final debiased contrastive loss of the user side can be written as follows:
\begin{equation}\label{mdfuser_eq}
\resizebox{.9\hsize}{!}{$\mathcal{L}_{fdbs\_user} = \sum_{u \in \mathcal{U}} -log \frac{exp(s(x_u^{'}, x_u^{''})/\tau + s(x_u^{'}, x_u^{M\_u})/\tau)}{\sum_{v\in \mathcal{U}}M_{user}(u,v) \cdot exp(s(x_u^{'},x_v^{''})/\tau)}$}
\end{equation}
% where $\gamma$ is the weight of potential positive samples. In the same way, we can acquire the final contrastive loss of the item side. So the formulation of the final self-supervised loss is as follows:
In the same way, we can acquire the final contrastive loss of the item side. So the formulation of the final self-supervised loss is as follows:
\begin{equation}\label{mdf_eq}
    \mathcal{L}_{fdbs\_ssl} = \mathcal{L}_{fdbs\_user} + \mathcal{L}_{fdbs\_item}
\end{equation}

\subsection{Multi-task learning}\label{3.4}
We follow the multi-task training strategy of SGL \cite{wu2021self} to jointly optimize the traditional recommendation tasks and the self-supervised learning tasks:
\begin{equation}\label{total_eq}
\mathcal{L}=\mathcal{L}_{rec}+\lambda \mathcal{L}_{fdbs\_ssl}+\mu\|\Theta\|_2^2
\end{equation}
where $\Theta$ denotes trainable parameters in recommendation tasks, and there are no additional parameters in our debiased contrastive learning method. $\lambda$ and $\mu$ are hyperparameters to control the proportion of self-supervised task and $L_{2}$ regularization $\Theta$, respectively.

\section{EXPERIMENTS}
To verify the effectiveness of the proposed TDSGL, we conduct extensive experiments and report detailed analysis results.
\subsection{Experimental Settings}
We conduct our experiments on three publicly available datasets: Yelp2018 \cite{he2020lightgcn,wu2021self}, Movilens-1M \cite{harper2015movielens} and LastFM \cite{cantador2011second}. Table \ref{tab:statistics} shows the statistics of the used datasets. In the training phase, we treat each observed user-item interaction as a positive instance, while the negative instance is typically generated by pairing the user with a random unobserved item.
\begin{table}[t]
\caption{\textbf{Statistics of the datasets.}}
\label{tab:statistics} % 设置标签
\setlength{\tabcolsep}{3.0mm}
{
\begin{tabular}{l|r|r|r|r}\hline
Dataset     & \#Users & \#Items & \#Interactions & \multicolumn{1}{l}{Sparisity} \\ \hline
Yelp2018    & 31668   & 38048   & 1,561,406      & 99.87\%                       \\
Movilens-1M & 6022    & 3043    & 895,699        & 95.11\%                       \\
LastFM      & 1891    & 15438   & 92,834         & 99.68\%                       \\ \hline
\end{tabular}
}
\end{table}
\paragraph{Baseline Methods}We compare TDSGL with the following five state-of-the-art methods to verify its superiority of performance,  covering MF-based methods (BPRMF \cite{rendle2012bpr}), GCN-based methods (NGCF \cite{wang2019neural}, LightGCN \cite{he2020lightgcn}, LR-GCCF \cite{chen2020revisiting}, DGCF \cite{wang2020disentangled}) and self-supervised methods (SGL’s \cite{wu2021self} variants: SGL-ED, SGL-ND, SGL-RW, where -ND denotes node dropout, -ED is short for edge dropout, and -RW means random walk). For fair comparisons, all the above methods are optimized by the same pairwise learning strategy. Each experiment in this section is conducted five times, we put great efforts to tune these methods based on the validation dataset and reported their best performance.
\paragraph{Evaluations method}As for the evaluation method, Recall@20 and NDCG@20 are chosen as the evaluation metrics as they are two widely used metrics in the evaluation of GCN-based CF models. Following \cite{he2020lightgcn,wu2021self}, we adopt the full-ranking strategy \cite{zhao2020revisiting}, which ranks all the candidate items that the user has not interacted with.
\subsection{Performance Comparison}
\begin{table}[t]
\caption{\textbf{Performance Comparison of Different Recommendation Models.}}
\label{tab:performance comparision} % 设置标签
\centering
\setlength{\tabcolsep}{1.8mm}
{
\begin{tabular}{c|cc|cc|cc}
\hline
Dataset  & \multicolumn{2}{c|}{Yelp2018}     & \multicolumn{2}{c|}{Movielens-1M} & \multicolumn{2}{c}{LastFM}        \\ \hline
Method   & Recall           & NDCG            & Recall          & NDCG            & Recall          & NDCG            \\ \hline
BPRMF    & 0.0307          & 0.0237          & 0.2389          & 0.2268          & 0.2262          & 0.2345          \\
NGCF     & 0.0555          & 0.0474          & 0.2361          & 0.2222          & 0.2470          & 0.2470          \\
DGCF     & 0.0640          & 0.0522          & 0.2620          & 0.2481          & 0.2382          & 0.2420          \\
LightGCN & 0.0649          & 0.0530          & 0.2419          & 0.2280          & 0.2417          & 0.2463          \\
LR-GCCF  & 0.0558          & 0.0343          & 0.2231          & 0.2124          & 0.2484          & 0.2540          \\ \hline
SGL-ND   & 0.0644          & 0.0528          & 0.2680          & 0.2527          & 0.2656          & 0.2809          \\
SGL-ED   & {\ul 0.0674}    & {\ul 0.0555}    & {\ul 0.2724}    & {\ul 0.2570}    & 0.2674          & 0.2827          \\
SGL-RW   & 0.0667          & 0.0547          & 0.2713          & 0.2559          & {\ul 0.2682}    & {\ul 0.2833}    \\ \hline
TDSGL     & \textbf{0.0695} & \textbf{0.0570} & \textbf{0.2739} & \textbf{0.2587} & \textbf{0.2752} & \textbf{0.2904} \\ \hline
\end{tabular}

The best result is \textbf{bolded} and the runner-up is \underline{underlined}.
}
\end{table}
Table \ref{tab:performance comparision} presents the comparison results of the overall performance, with the best and second-best results highlighted in bold and underlined, respectively. From the above results, we have several observations.
First, compared to BPRMF, a traditional CF-based method, graph-based collaborative filtering methods show better performance by exploring the high-order connectivity in the bipartite graph. Among all the graph collaborative filtering baseline models, LR-GCCF and LightGCN achieve the best performance in most cases, demonstrating the rationality and effectiveness of the simplified network architecture. However, some unexpected results can be found. We attributed the superior performance of DGCF on Movielens-1M to the sparser dataset, allowing the model to disentangle latent embeddings into multi-informative components. Additionally, NGCF performs worse than BPRMF on Movielens-1M, and we speculate that heavy GCN architecture may suffer from overfitting.

Second, all three SGL variants show more accurate recommendation results on all three datasets by incorporating a self-supervised task in the training process, reducing the influence of data sparsity and mitigating the over-smoothing issue. Among these variants, we find that SGL-ND is more unstable because dropping high-degree nodes will dramatically change the graph structure. Moreover, SGL-RW is equivalent to multi-layer SGL-ED, and these two models achieve the best performance in most cases. 

Finally, our proposed TDSGL consistently outperforms the baselines. We attribute this performance improvement to our debiased contrastive learning method in the self-supervised task, which can successfully decrease false negatives and explore the potential positive samples. From the observation, we speculate that the better performance on Yelp2018 and LastFM than Movielens-1M mainly results from fewer interactions. The topology-aware false negatives exploration method in the denser Movielens-1M is more challenging to recognize the false negatives because each pair of users (items) is more similar than that in sparse datasets.

\subsection{Further Analysis of TDSGL}
\paragraph{TDSGL vs. Negative mining techniques}

\begin{table}[t]
\centering
\caption{\textbf{Performance comparison of different false negative samples exploring methods.}} % 添加标题
\label{tab:Comparance with variants} % 设置标签
\setlength{\tabcolsep}{4.0mm}
% \resizebox{\textwidth}{15mm}
{
\begin{tabular}{ccccc}
\hline
\multicolumn{1}{l}{\multirow{2}{*}{Method}} & \multicolumn{2}{c}{Yelp2018}                                          & \multicolumn{2}{c}{ML-1M}                                            \\ \cline{2-5} 
\multicolumn{1}{l}{}                                 & \multicolumn{1}{l}{Recall} & \multicolumn{1}{l|}{NDCG} & \multicolumn{1}{l}{Recall} & \multicolumn{1}{l}{NDCG} \\ \hline
SGL                                                  & 0.0674                                 & \multicolumn{1}{c|}{0.0555}           & 0.2724                                 & 0.2570                               \\
SGL-PGCL                                             & 0.0676                                 & \multicolumn{1}{c|}{0.0556}           & 0.2187                                 & 0.2069                               \\
SGL-DGCL                                             & 0.0676                                 & \multicolumn{1}{c|}{0.0557}           & 0.2389                                 & 0.2278                               \\
SGL-SIM                                              & 0.0628                                 & \multicolumn{1}{c|}{0.0515}           & 0.2187                                 & 0.2069                               \\
TDSGL                                              & \textbf{0.0695}                                 & \multicolumn{1}{c|}{\textbf{0.0570}}         & \textbf{0.2739}                                 & \textbf{0.2587}                               \\ \hline
\end{tabular}}
\end{table}
In this section, we conduct false negatives exploration on feature space. Here we take several existing negative mining techniques covering SGL-SIM (false negatives exploration based on calculating embedding similarities), SGL-PGCL (false negatives exploration based on \cite{lin2022prototypical}), and SGL-DGCL (false negatives exploration based on \cite{zhao2021graph}) and report the results in Table \ref{tab:Comparance with variants}. From the table, we can observe that neither achieves the expected performance, in line with the findings of \cite{zhu2021empirical,xia2022progcl}, the existing false negatives exploration based on latent feature brings minor benefits to the GCL-based recommender system. 

\paragraph{Ablation study of TDSGL}
\begin{figure}[t]
\centering  %图片全局居中
\subfigure[Yelp]{
\label{Fig.sub.1}
\includegraphics[width=4.0cm,height = 2.2cm]{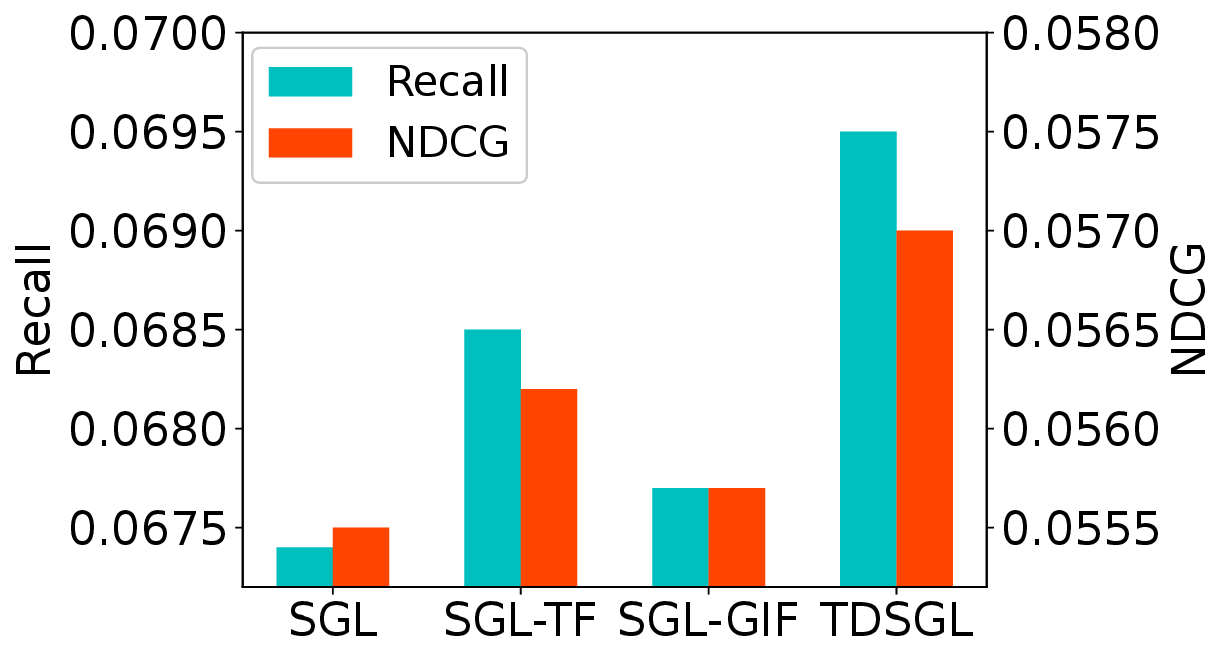}}
\subfigure[LastFM]{
\label{Fig.sub.2}
\includegraphics[width=4.0cm,height = 2.2cm]{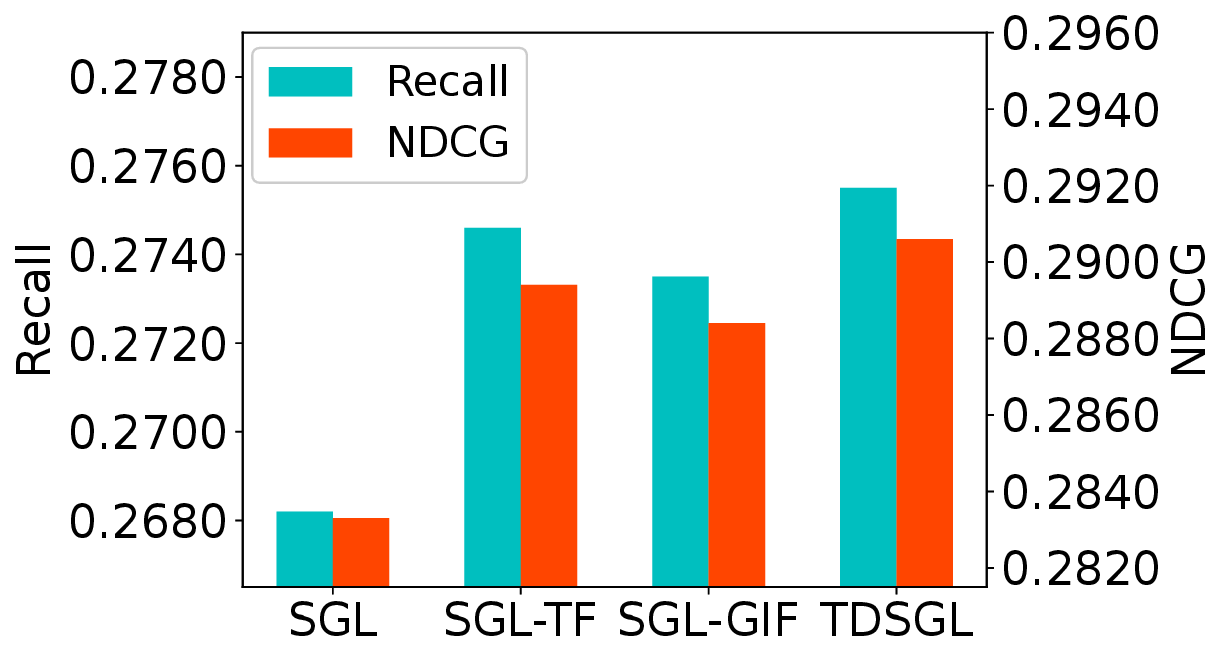}}
\caption{Performance comparison of variants of TDSGL}
\label{Fig 3}
\end{figure}
\begin{figure}[t]
\subfigure[Yelp]{
\label{Fig.sub.3}
\includegraphics[width=4.1cm,height = 2.2cm]{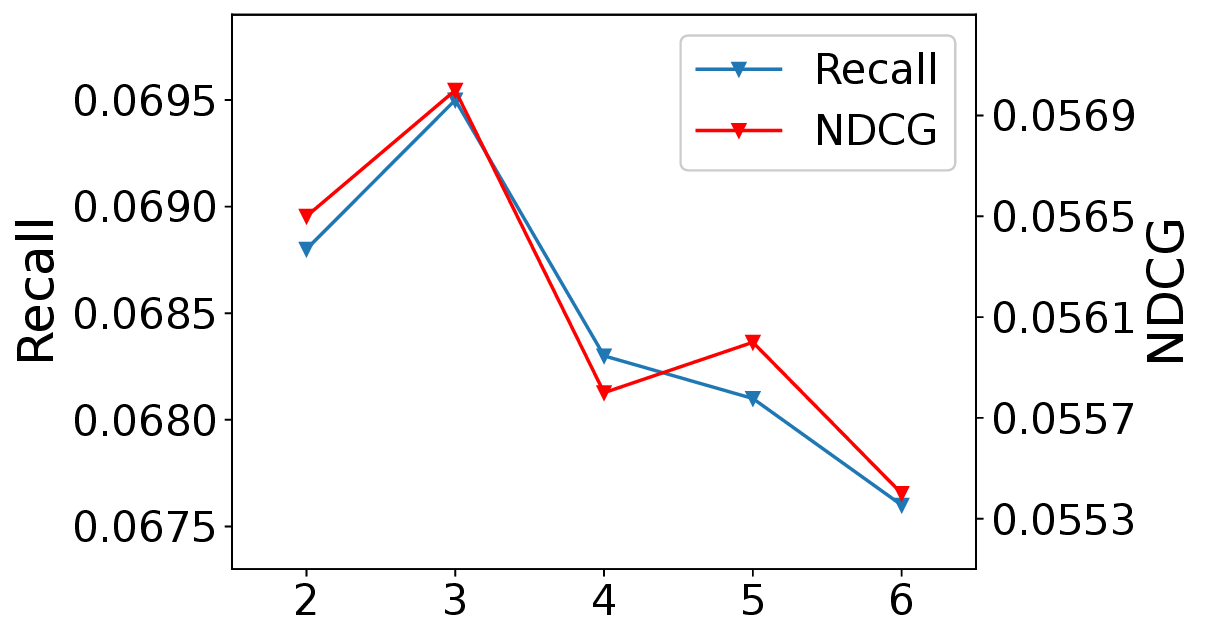}}
\subfigure[LastFM]{
\label{Fig.sub.4}
\includegraphics[width=4.1cm,height = 2.2cm]{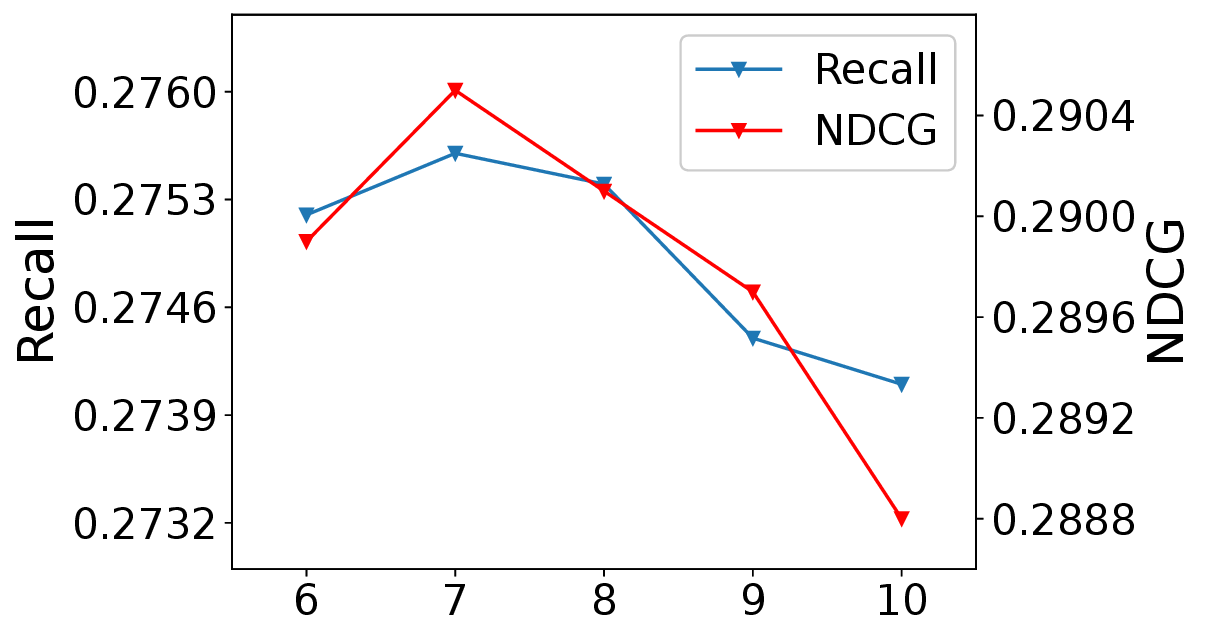}}
% \caption{Performance comparison of variants of TDSGL}
\caption{Impact of threshold $\beta$ of TDSGL.}
\label{Fig 4}
\end{figure}
Our proposed model TDSGL consists of two components: false negatives exploration and feature extraction module. In order to analyze the contributions of each component, we conduct an ablation study to analyze their contributions. We report the results in Fig. \ref{Fig.sub.1} and Fig. \ref{Fig.sub.2}, where the variants by removing the feature extraction module and false negatives exploration are denoted as -TF and -GIF, respectively. From the results, we can observe that removing each component in our model leads to performance degradation, while the two variants perform better than the baseline SGL. These results indicate that removing the false negative samples among negatives and exploring the potential positive samples improve the performance in GCL-based collaborative filtering.

\paragraph{Effect of Feature Extraction Module} 
\begin{table}[t]
\caption{\textbf{Components of the Feature Extraction Module}}
\label{tab:GCN variants} % 设置标签
\setlength{\tabcolsep}{2.0mm}
{
\begin{tabular}{c|cc|cc|cc}
\toprule 
\multirow{2}{*}{Model} & \multicolumn{2}{c|}{TDSGL(nl)} & \multicolumn{2}{c|}{TDSGL(nl + w)} & \multicolumn{2}{c}{TDSGL} \\ \cline{2-7} 
                       & Recall         & NDCG          & Recall          & NDCG           & Recall      & NDCG        \\ \hline
Yelp2018               & 00685          & 0.0563        & 0.0684          & 0.0562         & 0.0695      & 0.0570      \\
LastFM                 & 0.2757         & 0.2906        & 0.2742          & 0.2900         & 0.2756      & 0.2905      \\ 
\bottomrule
\end{tabular}
}
\end{table}
To investigate the impact of linear GCN in the feature extraction module, we replace it with nonlinear GCN (denoted as TDSGL(nl))  and nonlinear GCN with a transformation matrix (denoted as TDSGL(nl + w)), respectively, and report their effectiveness when incorporating them with contrastive learning in Table \ref{tab:GCN variants}. As observed, a linear GCN  outperforms other GCN variants on Yelp2018, which is in line with the founding in LightGCN that feature transformation and nonlinear activation impose a rather negative effect on the GCN module. Although the nonlinear GCN performs better than the linear one on LastFM, we argue that a simple linear GCN is more suitable for comprehensively capturing semantic information from false negatives.

\paragraph{Impact of Threshold $\beta$} 
 To analyze the influence of threshold $\beta$, we vary it in certain intervals, which are determined based on the characteristics of the datasets, and report the results in Fig. \ref{Fig.sub.3} and Fig. \ref{Fig.sub.4}. Specifically, for the more sparse Yelp2018 dataset, we tune the $\beta$ in a small interval[2, 3, 4, 5, 6]. From the results, we can find that when the $\beta$ increases from 2 to 6, the performance increases first and then decreases. This phenomenon is because when the $\beta$ is 2, the model mistakenly identifies many true negative samples as false ones. When the $\beta$ becomes large, the model will neglect many false negatives and degenerate to the SGL. On the more dense LastFM dataset, we also observe a similar phenomenon when varying $\beta$ within the range of [6, 7, 8, 9, 10].

 \section{CONCLUSIONS}
In this paper, we propose a novel contrastive learning model named Topology-aware Debiased Self-supervised Learning for Recommendation (TDSGL), which aims to mitigate the sampling bias caused by the random sampling strategy and explore the potential positive samples in graph contrastive learning. Firstly, we explore the false negatives based on the original user-item interaction data and improve the recommendation accuracy by removing them. Secondly, to further leverage potential positive samples, we design a feature extraction module to capture the latent feature of false negatives and regard the feature as a positive sample. We conduct extensive experiments on three publicly available datasets to demonstrate the effectiveness of the proposed TDSGL. In future work, we plan to utilize topological and semantic information to find more reliable false negative samples. Besides, we will study the more effective method to combine the false negative samples exploration with the feature extraction module to improve the performance.

\end{document}